\documentclass[fleqn,twoside]{article}
 \usepackage{espcrc1}
 \usepackage{graphicx}

 \newcommand{\be}{\begin{equation}}
 \newcommand{\ee}{\end{equation}}
 \def\SumInt{\sum\hspace{-1.3em}\int\,}
 \def\sumint{\hbox{$\Sigma$}\hspace{-0.63em}\raise0.1em\hbox{$\int$}\,}
 \def\gsim{\mathrel{\rlap{\lower0.2em\hbox{$\sim$}}\raise0.2em\hbox{$>$}}}

 \title{Resummation of the QCD thermodynamic potential\footnote{Talk
     given at the International Conference on Statistical QCD, Bielefeld,
     August 26-30, 2001.
     This work is supported by DOE under grant DE-AC02-98CH10886, and
     by the A.-v.-Humboldt foundation (Feodor-Lynen program).}}

\author{Andr\'e Peshier\address{%
        Brookhaven National Laboratory,
        Physics Department, Nuclear Theory,
        Upton, NY 11973, USA}}

\begin{document}

\maketitle

\begin{abstract}
  It is argued why thermodynamic approximations in terms of dressed
  propagators may, at larger coupling strength, be better behaved
  than perturbative results, and why in hot QCD the hard thermal loop
  approximation of the thermodynamic potential cannot be expected to
  work close to the phase transition.
\end{abstract}

\section{Introduction}
The thermodynamic potential is a quantity of central importance in
statistical field theory, and a lot of interest has been devoted to
its calculation. At high temperature, the perturbative expansion of
the thermodynamic potential is known to 5th order in the coupling for
the $\phi^4$ theory \cite{ParwaS}, \cite{BraatN4}, for QED
\cite{ParwaC}, and for QCD \cite{ZhaiK}, \cite{BraatN2}.
However, for physically relevant questions the applicability of these
results is limited; they are meaningful only when the coupling is so
weak that the thermodynamic potential is very close to its free limit.
Already for moderate values of the coupling the accuracy of the
approximation does not improve by the inclusion of higher order terms.

This feature of the perturbative expansion is similar to the behavior
of asymptotic series. Consider, e.\,g., the integral
$ Z(g^2) = \int_{-\infty}^\infty\! dx \exp\{ -L \} $, where $L =
\frac12\, x^2 + g^2x^4 $, which may be seen as a caricature of the
partition function ${\cal Z}$ in the $\phi^4$ theory \cite{ItzykZ}.
$Z(g^2)$ can be expressed in terms of a modified Bessel function. In
the spirit of perturbation theory, however, one would expand the
quartic `interaction part' of the exponential, and perform the
integrals before the sum to obtain an expansion in powers of $g^2$.
Being related to the number of diagrams in the perturbative calculation
of the partition function in the $\phi^4$ field theory, the
coefficients of this series,
$(-1)^n/n!\, \int\! dx\, x^{4n}\exp\{-x^2/2\} \sim (-1)^n n!$,
alternate in sign and grow asymptotically like the factorial of the
order $n$. Therefore, by the ratio test, the radius of convergence is
zero. Nonetheless, the coefficients contain relevant information. In
fact, the accuracy of the approximation of $Z(g^2)$ by the truncated
series does improve by higher-order terms as long as the truncation
order is smaller than an optimal $n^\star \sim 1/g^2$.

The divergence of the perturbative expansion of $Z(g^2)$ is obvious
already in its definition: in the complex $g^2$ plane, $Z$ has a cut
along the negative axis. Due to this essential feature one may presume
that some of the properties of $Z(g^2)$ and its expansion translate, at
least qualitatively, to path integrals and, hence, to the thermodynamic
potential $\Omega = -T\ln{\cal Z}$ in field theories. For example, the
analysis of the (non-)\,convergence of the perturbative expansion of
$\Omega$ demonstrates how the optimal truncation order decreases with
increasing coupling.
It is noted that for QCD close to the transition temperature already
the next-to leading order perturbative contribution cannot be
reconciled with the lattice data.

To motivate the following calculation of the thermodynamic potential
in terms of the propagator, I want to point out another feature of
the above toy model. Firstly, one can realize that the analog of the
propagator, $d(g^2) = \langle x^2 \rangle / Z$, where the weight of
the average is $\exp\{-L\}$, has also a cut along the negative $g^2$
axis. The same holds for the the `self-energy' $p(g^2) = d^{-1}(g^2)
-d^{-1}(0)$.
This suggests to consider $Z$ as a complex function of $p$. It turns
out that the mapping $p \rightarrow g^2 \rightarrow Z$ unfolds the
discontinuity along the negative $g^2$ axis. This analytic aspect is a
formal motivation to choose, in the physics context, a representation
of the thermodynamic potential in terms of the self-energy, or the
dressed propagator, as a starting point to derive approximations.
This is, of course, also supported by physical intuition. Dressed
Greens functions are expected to be a preferable basis for
nonperturbative approximations, since they resum entire subsets of
`bare' graphs. In particular, as known in other fields of physics, the
excitations described by the full propagator may be considered as
quasiparticles whose effective properties encode already a part of the
interaction effects. The idea to understand the thermodynamics of a
strongly coupled system in terms of its quasiparticles was
successfully demonstrated for hot QCD in phenomenological models where
the self-energies are approximated by their asymptotic values, see,
e.\,g., \cite{PeshiKPS3}.

In the following, a more systematic approach towards a quasiparticle
description of the thermodynamics of field theories is given, starting
from $\Phi$-derivable approximations. To outline this concept, the
scalar theory is considered first in the next Section, before
then turning to gauge theories.

\section{\boldmath$\Phi$-derivable approximations, scalar theory}
The thermodynamic potential was first expressed as a functional of the
exact propagator for a nonrelativistic fermion system \cite{LuttiW}.
This representation is the starting point for the so-called conserving
approximation scheme \cite{Baym}, in which diagrams are resummed in
thermodynamically consistent subsets.
In the scalar theory, the thermodynamic potential can be expressed by
the functional
\be
 \Omega[\Delta]
 =
 \frac12\, \SumInt\left[\ln(-\Delta^{-1})+\Delta\Pi\right]
 -\Phi[\Delta] \, ,
 \label{eq:Omega_4}
\ee%
evaluated at its stationary point, $\delta\Omega / \delta\Delta = 0$.
This condition is equivalent to $\Pi/2 = \delta\Phi / \delta\Delta$.
It defines the full propagator $\Delta$, which is related to the
self-energy $\Pi$ by Dyson's equation, $\Delta^{-1} = \Delta_0^{-1} -
\Pi$. $\Phi$ is given by the sum of the dressed 2-particle irreducible
bubble diagrams; for the interaction $g_0^2/4!\, \phi^4$,
\be
  \Phi =
  3\; \raise-2mm\hbox{\includegraphics[scale=0.55]{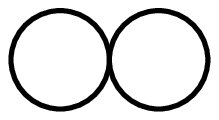}}
  + 12\;\raise-3mm\hbox{\includegraphics[scale=0.35]{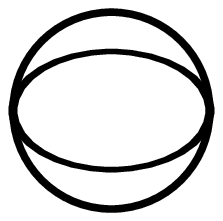}}
  + \ldots \, ,
  \label{eq:Phi_4}
\ee%
from which the expansion of $\Pi/2$ is obtained diagrammatically by
cutting one line in each graph. The loop truncation of these skeleton
expansions yields consistent (`$\Phi$-derivable') approximations
\cite{Baym}.

At leading-loop order, the self-energy is just a mass
term\footnote{Therefore, the resulting approximation is similar to
  that in screened perturbation theory \cite{KarscPP} with the
  `appropriate' choice for the mass parameter, introduced there to
  reorganize the perturbative expansion.}.
The $\Phi$-derivable approximation can be obtained explicitly
\cite{PeshiKPS4}, and it is equivalent to the large-$N$ limit of the
$O(N)$-symmetric scalar theory \cite{DrummHLR}. Considering, for
simplicity, the massless theory, and working in $3-2\epsilon$
dimensions in the $\overline{MS}$ scheme, the self-energy is to be
determined selfconsistently by
\be
 \Pi_{ll}
  =
  12\; \raise-1.5mm\hbox{\includegraphics[scale=0.55]{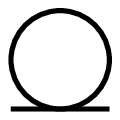}}\;
  =
  \frac{g_0^2}2
  \left(
    -\frac{\Pi_{ll}}{16\pi^2}
      \left[ \frac1\epsilon-\ln\frac{\Pi_{ll}}{\bar\mu^2}+1 \right]
    +\int_{k^3} \frac{n_b(\omega/T)}\omega
  \right) \, ,
  \quad
  \mbox{where} \quad
  \omega=(k^2+\Pi_{ll})^{1/2} \, .
  \label{eq:Pi_4 a}
\ee
The right hand side contains a divergent contribution $\sim \Pi_{ll}
/\epsilon$. This term, which implicitly depends on the temperature $T$,
is the vacuum part of the resummed quantum fluctuations, while the
medium contribution is finite due to the thermal distribution function
$n_b$. The divergence is absorbed in the bare coupling constant.
The renormalized coupling $g$ is defined as the scattering amplitude
at a momentum scale $s$. Resumming the chains\footnote{This yields
  only $1/3$ to the first coefficient of the $\beta$ function, which
  is an artefact of the truncation of the loop expansion. The two
  crossed scattering diagrams have to be omitted here since they would
  induce graphs beyond the superdaisy subset. This is consistent with
  the fact that they are suppressed in the $O(N\rightarrow\infty)$
  theory, where the leading-loop approximation becomes exact.},
$\includegraphics[scale=0.35]{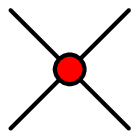} \rule[-2mm]{0mm}{6.5mm}
 = \includegraphics[scale=0.35]{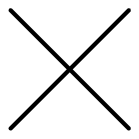}
  + 12\includegraphics[scale=0.35]{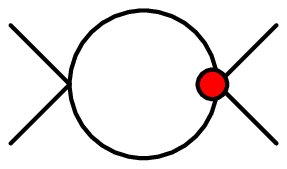}$,
yields
$g^2(s)
 =
 g_0^2-\frac12g_0^2 g^2(s) [1/\epsilon-\ln(-s/\bar\mu^2)+2]/(16\pi^2)$.
Expressing now $g_0$ by $g$ renders eq.\ (\ref{eq:Pi_4 a}) finite and
independent of $\bar\mu$,
\be
  \Pi_{ll}
  =
  \frac{g^2}2 \left(\;
    \frac{\Pi_{ll}}{16\pi^2}\left[ \ln\frac{\Pi_{ll}}{T^2} + 1 \right]
   +\int_{k^3} \frac{n_b(\omega/T)}\omega \right) ,
  \label{eq:Pi_4 b}
\ee%
where the coupling $g$ has been defined at $-s=T^2$. This gap equation
has a nontrivial structure. For $g$ smaller than $\tilde g \approx 10$,
there are two solutions. The smaller one is related to the perturbative
result which, by expanding the right hand side in $\Pi_{ll}/T^2$, is
reproduced up to order $g^3$. The second solution, related to the
tachyon \cite{DrummHLR}, is for small $g^2$ exponentially large and
thus of no physical relevance. As $g \rightarrow \tilde g$, the two
solutions approach the same value, $\tilde\Pi = \Pi_{ll}(\tilde g)
\approx 4T^2$, and the leading-loop approximation can no longer be
justified. For even larger values of the coupling, (\ref{eq:Pi_4 b})
has no solution.

At leading-loop order, the $\Phi$-contribution $\Phi_{ll} = \frac14\,
\sumint\Pi_{ll}\Delta_{ll}$ combines with the first term in
eq.~(\ref{eq:Omega_4}),
\be
 \Omega_{ll}
 =
  \frac12\,
  \SumInt\left[
     \ln(-\Delta_{ll}^{-1})
    +\textstyle\frac12\displaystyle\,\Delta_{ll}\Pi_{ll}
   \right]
 =
  \int_{k^3} T\ln\left( 1-\exp\{-\omega/T\} \right)
 -\frac{\Pi_{ll}}4\, \int_{k^3} \frac{n_b(\omega/T)}\omega
 -\frac{\Pi_{ll}^2}{128\pi^2} \, .
 \label{eq:Omega_4 ll}
\ee
I emphasize the relative factor of $\frac12$ between the two terms
under the sum-integral; it ensures that the divergent $T$-dependent
contributions of these terms cancel: $(\Pi_{ll}/8\pi)^2[ -1/\epsilon +
\frac12\,2/\epsilon] = 0$.
The first integral in eq.\ (\ref{eq:Omega_4 ll}) is the thermodynamic
potential of a system of quasiparticles with mass $\Pi_{ll}^{1/2}$.
Expanded in $\Pi_{ll}/T^2 \sim g^2$, it over-includes the perturbative
${\cal O}(g^2)$ term by a factor of two.
This is compensated by the second integral -- in fact, the
approximation (\ref{eq:Omega_4 ll}) reproduces the ${\cal O}(g^3)$
result. The term $\sim \Pi_{ll}^2$ is a remnant of the resummed vacuum
fluctuations. It leads to a minimum in the pressure, $p = -\Omega/V$,
considered as a function of the self-energy, see Fig.~1. Due to the
stationary property of $\Omega_{ll}$, the location of this minimum
coincides with $\tilde\Pi_{ll}$, and hence indicates also where the
leading-loop approximation breaks down.

This observation is interesting. With regard to the next section,
assume we had solved the gap equation only perturbatively, to leading
order: $\Pi_{ll} \approx \Pi^\star = (g_0 T)^2/24$ (without
resummation, the coupling remains bare). This is equivalent to the
hard thermal loop (HTL) approximation of $\Pi$. Approximating now the
$\Phi$-contribution by $\Phi^\star = \frac14\,
\sumint\Pi^\star\Delta^\star$ yields the {\em same} functional form
(\ref{eq:Omega_4 ll}) for $\Omega^\star(\Pi^\star)$ as in the
selfconsistent approximation\footnote{Of course, it is not the same
  function of the coupling, and it can reproduce the perturbative
  result only up to leading order (as an aside, the next-to leading
  order term comes out too small by a factor of $\frac14$).  Still, it
  is a useful approximation since the quasiparticle mass is a
  measurable quantity.  Note also that an ansatz $\Phi^\star =
  3g_0^2(\sumint\Delta^\star)^2$ has to be abandoned since it leads to
  uncompensated $T$-dependent divergences in the thermodynamic
  potential. \protect\label{fn: a}}. Therefore, the maximal value of
the self-energy, and hence for the coupling, beyond which the
approximation cannot be meaningful (which could not be surmised from
$\Pi^\star = (g_0 T)^2/24$) can be inferred also from the HTL
approximation $\Omega^\star(\Pi^\star)$, which is important for a later
argument.

Another useful thermodynamic quantity is the entropy density, $s = dp /
dT$. In the leading-loop $\Phi$-derivable approximation, it can
alternatively be obtained from a 1-loop functional of the dressed
propagator \cite{VandeB}, \cite{BlaizIR3}, which is not specific to the
$\phi^4$ theory. In this case, the result is equivalent to the entropy
of a system of quasiparticles with mass $\Pi_{ll}^{1/2}$ in the
selfconsistent approximation, or $(\Pi^\star)^{1/2}$ in the HTL
approximation.  Therefore, the entropy is a monotonously decreasing
function of the quasiparticle mass, see Fig.~1. It shows, in contrast
to the pressure, no sign of the breakdown of the approximation. This
observation is consistent with the thermodynamic relation, since to
reconstruct $p$ from $s$ one needs the $T$-dependence of the
self-energy. It is also plausible on general grounds. While the
pressure and the self-energy contain vacuum contributions, which become
important at larger coupling, the entropy is determined only by thermal
excitations.

\section{Gauge theories, QCD}
Aside from the technical complications to solve a set of coupled
nonlocal Dyson equations selfconsistently, it may not be desired in
gauge theories. In the $\Phi$-derivable approximation scheme, `bare'
diagrams of all orders in the coupling are grouped in subsets to
preserve thermodynamic consistency. Since this procedure distinguishes
the 2-point functions in the hierarchy of Greens functions, it cannot
be expected to also preserve gauge invariance (unless solved exactly
to all orders in the loop expansion). Another, technical, issue is how
to accomplish the renormalization of resummed self-energies at finite
temperature.
Approximating the self-energies at leading order in the coupling makes
the renormalization trivial. This would correspond to the
considerations at the end of the Sec.\,2, and lead to similar
consequences. For example, only the leading order of the pressure could
be reproduced upon expanding the resulting expression, while the (gauge
dependent) higher order terms would not match the perturbative ones.

However, it turns out that approximating the self-energies by their HTL
contributions, which are gauge invariant, suffices to resum the leading
contributions. When calculating the thermodynamic potential, this
additional approximation requires to account for the
$\Phi$-contribution\footnote{This is not an issue in the entropy
    calculation \cite{BlaizIR3} because the leading loop
    entropy-functional is independent of $\Phi$.}
more carefully than in the $\phi^4$ theory. For simplicity of the
argument, consider first the case of QED,
\be\textstyle
  \Omega
   =
  \frac12\, {\rm Tr}\! \left[\, \ln(-D^{-1})+D\Pi\, \right]
  - {\rm Tr}\! \left[\, \ln(-S^{-1})+S\Sigma\, \right]
  - \Phi[D,S] \, ,
  \qquad
  \Pi = 2\, \delta\Phi / \delta D \, ,
  \quad
  \Sigma = -\delta\Phi / \delta S \, ,
  \label{Omega_QED}
\ee
where $D = (D_0^{-1}-\Pi)^{-1}$ is the photon propagator, and $S =
(S_0^{-1}-\Sigma)^{-1}$ is the electron propagator. The Lorentz and
spinor indices are implicit here, and summed over in the traces.
At leading loop order,
\be
  \Phi_{ll} = \frac12\includegraphics[scale=0.35]{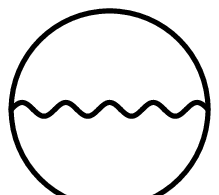}\, , \quad
  \Pi_{ll}  = \includegraphics[scale=0.35]{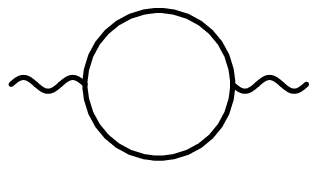} \, , \quad
  \Sigma_{ll} = - \includegraphics[scale=0.35]{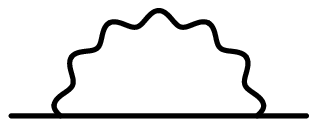} \, .
  \label{graphs_QED}
\ee
Both in the selfconsistent and in the perturbative approximation, the
$\Phi$-contribution to the thermodynamic potential could be evaluated
by closing the external legs\footnote{This is the same strategy as in
  the $\phi^4$ theory, see footnote \ref{fn: a}. Evaluating, instead,
  the functional $\Phi_{ll}$ with the perturbative propagators would
  also in the present case result in uncompensated $T$-dependent
  divergences.}
of, either, the photon or the electron self-energy, or a linear
combination of those traces.  This is not the case in the HTL
approximation, where terms of order of the external momentum squared
are neglected for the contributions $\sim T^2$. However, when traced
over all momenta, these terms contribute to the same order, $e^2T^4$,
as the HTL self-energies.
This does not indicate that the HTL approximation is inutile
here\footnote{It is noted that the leading contribution in the final
  result for the thermodynamic potential comes from large momenta
  close to the light cone, where the HTL approximation is justified.},
but rather suggests the solution of this riddle, namely by keeping
track of all relevant terms in $\Phi_{ll}$. At order $e^2$, $\Phi_{ll}$
is given by a double sum-integral over an expression with a numerator
$N = K^2-Q_1^2 -Q_2^2$, where $K$ is the photon momentum, and $Q_{1,2}$
are the electron momenta. One of the three terms in $N$ would be
neglected in the HTL approximation, e.\,g., closing the legs of the
photon HTL self-energy $\Pi^\star$ amounts to drop $K^2$. Summing over
the cases where one of the terms in $N$ is dropped yields $2N$. Hence,
$ \Phi^\star
  =
  \frac14\, {\rm Tr}\Pi^\star D^\star
  -\frac12\, {\rm Tr}\Sigma^\star S^\star
  \label{PhiStar_QED}
$
is equivalent to $\Phi_{ll}$ up to terms of higher order.
The resulting approximation \cite{Peshi2} for $\Omega$,
\be\textstyle
 \Omega^\star
 =
 \frac12\, {\rm Tr}\!
   \big[\, \ln(-D^{\star\, -1})+\frac12\,\Pi^\star D^\star\, \big]
 -{\rm Tr}\!
   \big[\, \ln(-S^{\star\, -1})+\frac12\,\Sigma^\star S_\star\, \big] ,
\ee
is in close analogy (notice the factors $\frac12$ in the trace terms)
to the expression in the scalar theory.

For QCD, a similar (though due to the non-Abelian topology slightly
more involved) analysis of the HTL $\Phi$-contribution was given in
\cite{Peshi2}. Let me here argue directly for a SU($N_c)$ gluon plasma
(including ghosts to compensate the unphysical degrees of freedom) why
to expect the expression
\be\textstyle
 \Omega_g^\star
  =
 \frac12\, {\rm Tr}_c\!
   \big[\, \ln(-D_g^{\star\, -1})+\frac12\,\Pi_g^\star D_g^\star\, \big]
 -{\rm Tr}_c\! \big[ \ln(-D_{ghost, 0}^{-1}) \big]\, .
 \label{eq: OmegaStar_QCD}
\ee%
The transverse and longitudinal components of the gluon HTL self-energy
$\Pi^\star(p_0,p)$ are given by dimensionless functions of $p_0/p$,
times the asymptotic gluon mass squared, $M_g^2 = \frac16\, N_c\,
g^2T^2$. Therefore, $M_g^2 \partial_{M_g^2} \ln(-D_g^{\star\,-1}) =
-\Pi_g^\star D_g^\star$.
Hence, the factor $\frac12$ between the two terms in the gluon trace
is indispensable to cancel their individually divergent contributions,
\begin{figure}[hb]
 \begin{minipage}[t]{75mm}
  \hskip-5mm
  \includegraphics[scale=0.72]{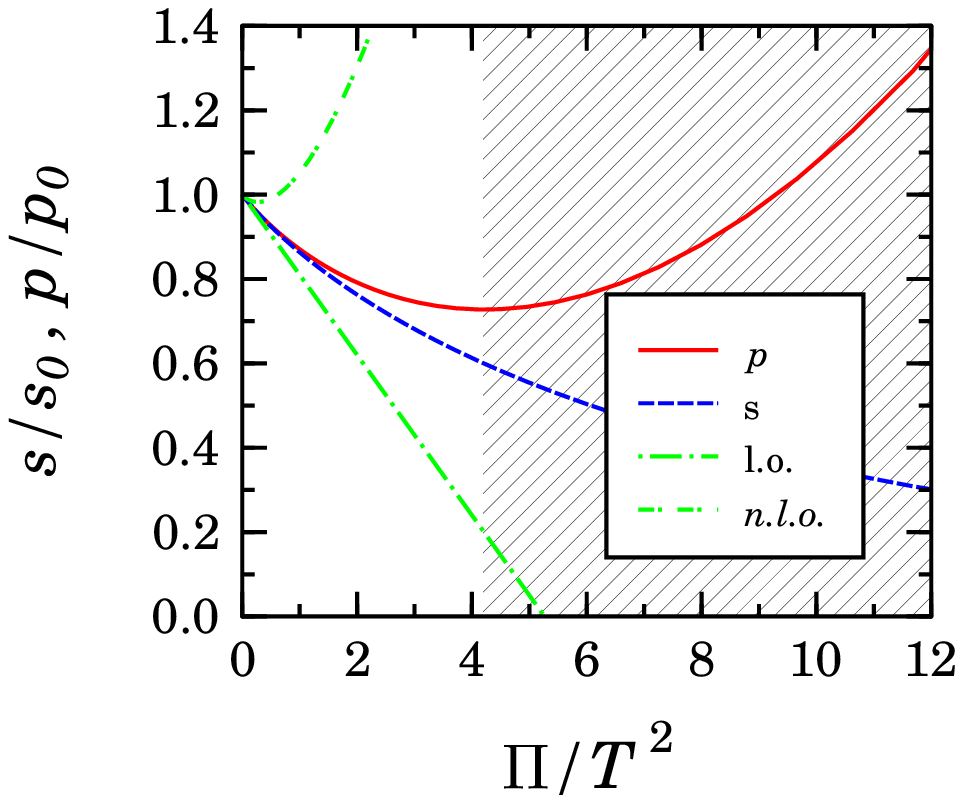}
 \end{minipage}
 \hspace{\fill}
 \begin{minipage}[t]{75mm}
  \hskip-5mm
  \includegraphics[scale=0.72]{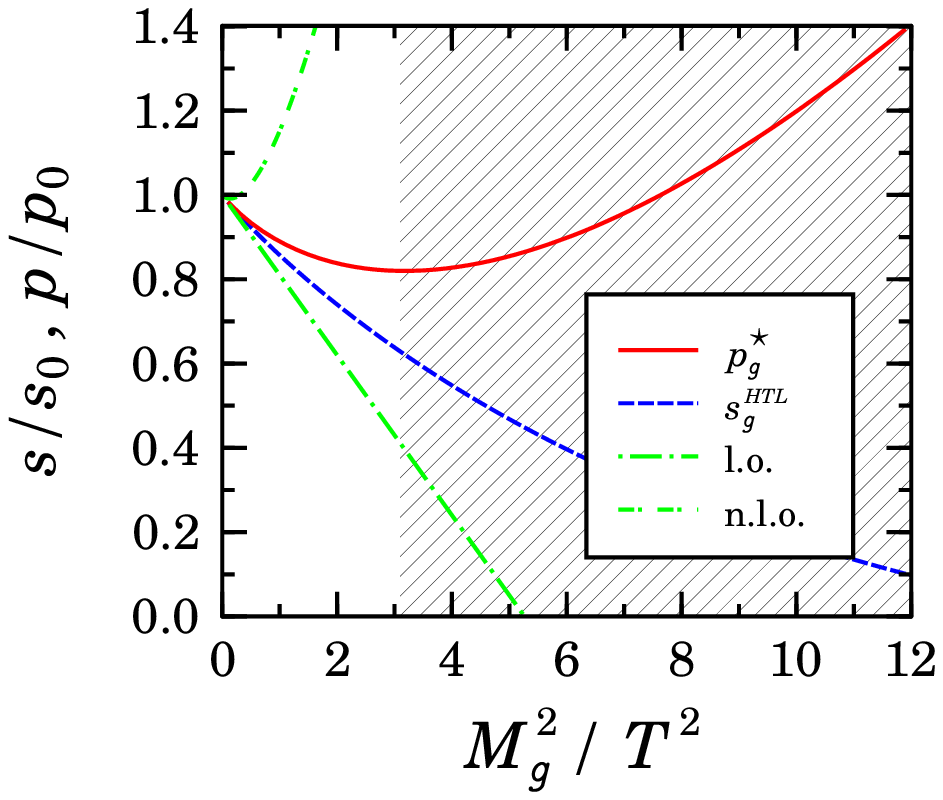}
 \end{minipage}
 \vskip-1cm
 \caption{Pressure and entropy in the HTL approximation, as functions
   of the thermal mass scale in the $\phi^4$ theory (left), and for
   SU($N_c$) (right).  Shown also are the leading and next-to leading
   perturbative results.  In the $\phi^4$ theory, the selfconsistent
   and the HTL approximation have the same functional dependence on
   the respective self-energy.  In the hatched region the
   approximations are not justified.}  \vskip-2mm\noindent
\end{figure}
which are $\sim M_g^4 / \epsilon^n$ on dimensional grounds. The result
(\ref{eq: OmegaStar_QCD}) has a striking similarity to the scalar case.
Its expansion reproduces the ${\cal O}(g^2)$ result, but
under-estimates the ${\cal O}(g^3)$ term by a factor of $\frac14$.
The pressure $p_g^\star$ has a minimum at $\tilde{M}_g^2 \approx 3T^2$
(see Fig.~1). Similar to the $\phi^4$ theory, the dip is caused by a
term $\sim M_g^4$ implicit in (\ref{eq: OmegaStar_QCD}). It originates
from the vacuum contributions, which for larger coupling become as
important as the thermal contributions.
This is in accordance with the general argumentation in Sec.~2, and not
specific to the HTL approximation. I therefore conjecture also for QCD,
where the solution structure of the selfconsistent Dyson equation
cannot be studied explicitly, that the leading-loop approximation may
not be meaningful for $M_g \gsim \tilde{M}_g$. Again, this could not be
inferred from the HTL entropy $s_g^{_{HTL}}$ \cite{BlaizIR3} which
decreases monotonously with $M_g^2$.
This conjecture is supported by lattice data.
\begin{figure}[h]
 \begin{minipage}[b]{75mm}
  \hskip-5mm
  \includegraphics[scale=0.72]{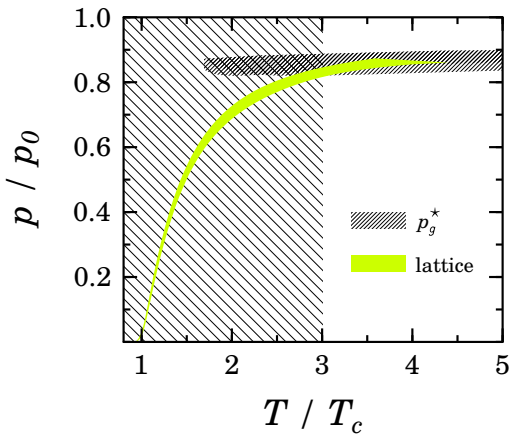}
  \vskip-2mm\noindent
 \end{minipage}
 \hspace{\fill}
 \begin{minipage}[b]{8cm}
   \caption{The pressure as a function of the temperature in the SU(3)
     theory. The shaded band shows the lattice data \cite{BoydEtAl},
     \cite{OkamoEtAl}. The hatched band for $p_g^\star$ represents the
     variation of the result when determining the temperature
     dependence of $M_g^2$ by, either, the 1- or the 2-loop running
     coupling in the $\overline{MS}$ scheme, assuming $T_c = 1.14
     \Lambda_{\overline{MS}}$, and varying the renormalization scale
     $1 \le \bar\mu/(\pi T) \le 4$ as in \cite{BlaizIR3}.  
     In accordance with the conjecture in the text (but not shown here), 
     also the HTL entropy $s_g^{_{HTL}}(T)$ starts to deviate clearly
     from the lattice data below $3T_c$ \cite{BlaizIR3}.}  
     \vskip-2mm\noindent
 \end{minipage}
\end{figure}
At $\tilde M_g$, $p_g^\star$ is approximately 80\% of the free
pressure, a value reached by the lattice data at $3T_c$ ($T_c$ is the
confinement temperature). At this temperature, indeed, $p_g^\star(T)$
starts to match the data, see Fig.~2, as does the entropy
$s_g^{_{HTL}}$ \cite{BlaizIR3}.

In conclusion, it was shown that the HTL approximations of
thermodynamic observables has a considerably improved range of
applicability in the large coupling regime compared to perturbative
results. For hot QCD, the approximation works down to temperatures 
of a few times above $T_c$. This encourages developments to extend,
similar to the approach \cite{PeshiKS1}, the formalism to finite
chemical potential, where no lattice data are available so far.

\end{document}